\documentclass[12pt]{article}

\usepackage[margin=1in]{geometry}
\usepackage{setspace}
\usepackage{times}
\usepackage{amsmath}
\usepackage{graphicx}
\usepackage{booktabs}
\usepackage{float}
\usepackage{natbib}
\usepackage{hyperref}
\usepackage{tikz}
\usepackage{caption}
\usepackage{tabularx}
\usepackage[table]{xcolor}
\usepackage{array}
\usepackage{ragged2e}
\usepackage{makecell}
\newcolumntype{Y}{>{\RaggedRight\arraybackslash}X}

\title{Technical Report: Causal ANOVA Applied to COMPAS \\ 
\large A Causal Analysis of Demographic Explainability in Recidivism Prediction}
\author{Yiming Xiong and Zijun Gao}
\date{\today}

\begin{document}

\maketitle

% --------------------
% Abstract
% --------------------
\begin{abstract}
This report applies Causal ANOVA to the COMPAS recidivism prediction dataset to study how demographic variables contribute to variation in two-year recidivism outcomes. Using a directed acyclic graph (DAG) to encode causal assumptions, we estimate counterfactual explainability scores for Race, Sex, and Age Category. We then examine whether the explanatory roles of Race and Sex change after conditioning on age group. The results suggest that Age Category explains the largest share of variation in the full causal model. After conditioning on age, Sex explainability is highest among younger defendants and declines with age, while Race explainability increases with age and is most pronounced among older defendants. As a complementary analysis, we use logistic regression models to interpret the demographic patterns suggested by the Causal ANOVA results.
We provide a Python package of Causal ANOVA at \url{https://github.com/devonxiong/causal-anova}.
\end{abstract}


\newpage
\tableofcontents
\newpage

% =====================================================
\section{Dataset and Causal Framework}
% =====================================================

\subsection{The COMPAS Dataset}

The COMPAS (Correctional Offender Management Profiling for Alternative Sanctions) dataset contains criminal justice records used to predict the likelihood of recidivism. The dataset includes demographic attributes such as Race, Sex, and Age Category, as well as criminal history variables such as Priors Count and Charge Degree. The outcome variable is Two-Year Recidivism, which records whether a defendant reoffended within two years.

In this report, we focus on the explanatory role of three demographic variables: Race, Sex, and Age Category. These variables are treated as root nodes in the causal graph. Priors Count and Charge Degree are treated as mediator variables because they may transmit part of the association between demographic attributes and the final recidivism outcome.

\subsection{Causal DAG}

To encode the causal assumptions formally, we construct a directed acyclic graph in which Race, Sex, and Age Category are treated as root nodes. These demographic variables are assumed to influence Priors Count and Charge Degree, which serve as mediator variables in the causal graph. The outcome node, Two-Year Recidivism, is then affected by the demographic root nodes and by the mediator nodes.

This structure allows demographic variables to affect the outcome in two ways. First, they may have direct paths to the outcome. Second, they may affect the outcome indirectly through Priors Count and Charge Degree. The purpose of this DAG is not to claim that the graph captures every possible social or legal mechanism, but to provide a formal structure for estimating counterfactual explainability scores.

\begin{figure}[H]
\centering
\begin{tikzpicture}[
    >=stealth,
    every node/.style={font=\normalfont\normalsize}
]
    % Nodes
    \node (race)   at (0,  2.2) {Race};
    \node (sex)    at (0,  0.0) {Sex};
    \node (age)    at (0, -2.2) {Age Category};

    \node (priors) at (4.2, 1.2) {Priors Count};
    \node (charge) at (4.2,-1.0) {Charge Degree};

    \node (recid)  at (8.3, 0.1) {Two-Year Recidivism};

    % Arrows from Race
    \draw[->] (race) -- (priors);
    \draw[->] (race) -- (charge);
    \draw[->] (race) to[bend left=8] (recid);

    % Arrows from Sex
    \draw[->] (sex) -- (priors);
    \draw[->] (sex) -- (charge);
    \draw[->] (sex) -- (recid);

    % Arrows from Age Category
    \draw[->] (age) -- (priors);
    \draw[->] (age) -- (charge);
    \draw[->] (age) to[bend right=8] (recid);

    % Internal arrows
    \draw[->] (priors) -- (charge);
    \draw[->] (priors) -- (recid);
    \draw[->] (charge) -- (recid);
\end{tikzpicture}
\caption{DAG for COMPAS causal explainability analysis.}
\label{fig:compas_dag}
\end{figure}
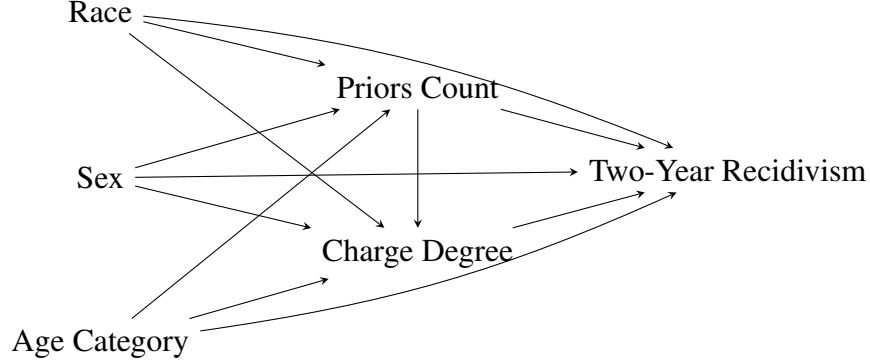

% =====================================================
\section{Overall Counterfactual Explainability}
% =====================================================

\subsection{Counterfactual Explainability Method}
To estimate counterfactual explainability, we use a Pick-Freeze simulation procedure of \cite{gao2025counterfactual}. First, we generate a factual world by sampling independent noise variables for each node in the DAG. Then, for a target subset of variables, we replace the noise terms for those variables with an independent set of noise values while keeping the remaining variables fixed. This produces a counterfactual version of the outcome. Finally, we compare the variance between the factual and counterfactual outcomes to the variance between two fully independent outcome simulations.
The explainability score is computed as
\[
\xi_G\!\left( \bigvee_{k\in S} V_k \Rightarrow Y \right)
=
\frac{\operatorname{Var}\!\big( f(V^{*}) - f(V^{*\prime}_{S},\, V^{*}_{-S}) \big)}
     {2\,\operatorname{Var}\!\big( f(V^{*}) \big)}.
\]
where $S$ is the subset of variables being explained, $f(V^{*})$ is the factual simulated outcome, $f(V^{*\prime}_{S},\, V^{*}_{-S})$ is the counterfactual outcome after replacing the noise for variables in $S$, and $V^{*\prime}$ is an independent draw of all noise variables.
Intuitively, a larger value of $\xi_G\!\left( \bigvee_{k\in S} V_k \Rightarrow Y \right)$ means that changing the variables in $S$ produces a larger change in the simulated outcome. Therefore, $\xi_G\!\left( \bigvee_{k\in S} V_k \Rightarrow Y \right)$ measures how much explanatory power the subset $S$ has under the assumed causal structure.

\subsection{Estimation}

The explainability scores are estimated as follows. Root nodes (Sex, Race, and Age Category) in the full DAG are sampled directly from their empirical distributions. For non-root nodes, we estimate conditional quantile functions $Q_k(\cdot \mid V_{\mathrm{pa}(k)})$ by gradient boosting with quantile loss on a grid $\{0.01, 0.03, \dots, 0.99\}$, apply isotonic rearrangement to enforce monotonicity, and recover values by piecewise-linear interpolation between grid points with clamping outside the grid. To construct the Pick-Freeze estimator, we draw independent uniform noise $E = (E_1, \dots, E_{K+1})$ and an independent copy $E'$, then generate outcomes sequentially via $V_k = Q_k(E_k \mid V_{\mathrm{pa}(k)})$, yielding a factual outcome $f(E)$, a counterfactual outcome $f(E'_S, E_{-S})$ in which the noise of subset $S$ is resampled while $E_{-S}$ is frozen, and an independent outcome $f(E')$. The total explainability $\xi_G(\bigvee_{k\in S} V_k \Rightarrow Y)$ is then estimated as the ratio of sample variances $\mathrm{Var}(f(E) - f(E'_S, E_{-S}))$ over $\mathrm{Var}(f(E) - f(E'))$. Within each repetition, the same draws $E$ and $E'$ are shared across all subsets $S$ via common random numbers (CRN), so the contrasts entering the inclusion-exclusion formula are coupled; this coupling keeps interaction terms non-negative and ensures that higher-order interactions remain below their constituent lower-order terms. We use $N = 10{,}000$ Monte Carlo samples per repetition, average over $R = 8$ repetitions, and report the mean and standard error. Interaction terms $\xi_G(\bigwedge_{k\in S} V_k \Rightarrow Y)$ and all other clauses in $\mathcal{E}(V)$ are obtained from the total explainabilities via inclusion-exclusion.

\subsection{Total Counterfactual Explainability}

Table~\ref{tab:full_explainability_scores} presents the counterfactual explainability score $\xi$ for all subsets of the demographic root nodes under the full DAG in Figure~\ref{fig:compas_dag}. The explainability score quantifies how much variation in the outcome is explained by a given set of variables under the assumed causal structure.

Among the three individual root nodes, Age Category has the highest explainability score ($\xi = 0.1283$), followed by Race ($\xi = 0.0497$) and Sex ($\xi = 0.0336$). This suggests that age-related variation explains the largest share of variation in two-year recidivism under the full causal model. The joint effect of all three variables reaches $\xi = 0.1659$, which indicates that the demographic root nodes together explain a larger portion of the outcome variation than any single demographic variable alone.

\begin{table}[H]
    \centering
    \small
    \setlength{\tabcolsep}{4pt}
    \renewcommand{\arraystretch}{0.9}
    \caption{Counterfactual explainability of COMPAS root nodes.}
    \label{tab:full_explainability_scores}
    \begingroup
    \small
    \setstretch{1}
    \setlength{\tabcolsep}{3pt}
    \renewcommand{\arraystretch}{0.85}
    \setlength{\aboverulesep}{2pt}
    \setlength{\belowrulesep}{2pt}
    \begin{tabular}{@{}lc@{\hspace{1.5em}}lc@{}}
        \toprule
        \textbf{Variables} & $\boldsymbol{\xi}$ &
        \textbf{Variables} & $\boldsymbol{\xi}$ \\
        \midrule
        Sex & 0.0336
            & Sex, Race, Age Category & 0.1659 \\
        Race & 0.0497
            & Sex $\wedge$ Race & 0.0114 \\
        Age Category & 0.1283
            & Sex $\wedge$ Age Category & 0.0166 \\
        Sex, Race & 0.0719
            & Race $\wedge$ Age Category & 0.0229 \\
        Sex, Age Category & 0.1453
            & Sex $\wedge$ Race $\wedge$ Age Category & 0.0053 \\
        Race, Age Category & 0.1551
            & & \\
        \bottomrule
    \end{tabular}
    \endgroup
\end{table}

\subsection{Explainability Interaction}

The interaction terms provide additional information beyond the marginal explainability scores. A positive interaction score means that the joint explanatory contribution of two or more variables is not fully captured by their separate marginal contributions. In this analysis, the Sex $\wedge$ Race interaction score is $\xi = 0.0114$, suggesting that Sex and Race do not contribute to outcome variation in a purely additive way. The three-way interaction among Sex, Race, and Age Category has a score of $\xi = 0.0053$, which is smaller than each of the two-way interactions, consistent with the hierarchy that higher-order interactions capture progressively smaller shares of the outcome variation.

This interaction pattern motivates the age-conditioned analysis in the next section. A non-zero three-way interaction means that the joint explanatory relationship between Sex and Race varies across age groups. We therefore condition on age group to examine how the explanatory roles of Sex and Race, and their interaction, change across age strata.

\subsection{Logistic Regression as a Complementary Interpretation}

As a complementary analysis, we run a logistic regression model predicting two-year recidivism from Sex, Race, Age Category, and all two-way and three-way interaction terms among them. This model is fit on the real COMPAS dataset. It is not meant to replace the Causal ANOVA framework; rather, it provides a coefficient-based interpretation of the direction and relative magnitude of demographic effects suggested by the explainability scores. Consistent with the recommendation in \cite{gao2025counterfactual}, we exclude mediator variables such as Priors Count and Charge Degree from this regression, because conditioning on a descendant would block causal pathways such as Race $\to$ Priors Count $\to$ Two-Year Recidivism and would therefore understate the total explanatory role of the root nodes.

To identify the most informative reference category, we fit the saturated logistic regression under all 24 possible combinations of reference categories for Sex, Race, and Age Category, and selected the combination yielding the largest number of statistically significant coefficients. The optimal reference group is Male, African-American, and 25-45 years of age, yielding 8 significant coefficients and reflecting the highest recidivism subgroup in the data. Table~\ref{tab:logit_af_ref} presents the results under this reference category.

\begin{table}[tbp]
    \centering
    \caption{Logistic Regression: Two-Year Recidivism on Demographics (Reference: Male, African-American, Age 25--45)}
    \label{tab:logit_af_ref}
    \small
    \begingroup
    \small
    \setstretch{1}
    \setlength{\tabcolsep}{3pt}
    \renewcommand{\arraystretch}{0.85}
    \setlength{\aboverulesep}{2pt}
    \setlength{\belowrulesep}{2pt}
    \begin{tabular}{lcccc}
        \toprule
        \textbf{Predictor} & \textbf{Coef.} & \textbf{Std.\ Err.} & \textbf{$z$} & \textbf{$p$-value} \\
        \midrule
        \multicolumn{5}{l}{\textit{Main Effects}} \\
        Intercept & 0.1325 & 0.047 & 2.804 & 0.005 \\
        Sex: Female & $-$0.6124 & 0.114 & $-$5.380 & $<$0.001 \\
        Race: Caucasian & $-$0.3993 & 0.079 & $-$5.033 & $<$0.001 \\
        Race: Hispanic & $-$0.6433 & 0.128 & $-$5.044 & $<$0.001 \\
        Race: Other & $-$0.7876 & 0.158 & $-$4.990 & $<$0.001 \\
        Age: Greater than 45 & $-$0.4344 & 0.103 & $-$4.236 & $<$0.001 \\
        Age: Less than 25 & 0.4682 & 0.090 & 5.216 & $<$0.001 \\
        \midrule
        \multicolumn{5}{l}{\textit{Two-Way Interactions}} \\
        Female $\times$ Caucasian & 0.5726 & 0.174 & 3.291 & 0.001 \\
        Female $\times$ Hispanic & 0.1425 & 0.321 & 0.445 & 0.657 \\
        Female $\times$ Other & 0.2981 & 0.401 & 0.743 & 0.458 \\
        Female $\times$ Age $>$ 45 & $-$0.3095 & 0.293 & $-$1.056 & 0.291 \\
        Female $\times$ Age $<$ 25 & $-$0.1901 & 0.207 & $-$0.920 & 0.357 \\
        Caucasian $\times$ Age $>$ 45 & $-$0.1984 & 0.151 & $-$1.309 & 0.190 \\
        Hispanic $\times$ Age $>$ 45 & $-$0.3400 & 0.271 & $-$1.253 & 0.210 \\
        Other $\times$ Age $>$ 45 & 0.1645 & 0.306 & 0.537 & 0.591 \\
        Caucasian $\times$ Age $<$ 25 & $-$0.0360 & 0.159 & $-$0.226 & 0.821 \\
        Hispanic $\times$ Age $<$ 25 & 0.1884 & 0.242 & 0.778 & 0.437 \\
        Other $\times$ Age $<$ 25 & 0.4218 & 0.289 & 1.461 & 0.144 \\
        \midrule
        \multicolumn{5}{l}{\textit{Three-Way Interactions}} \\
        Female $\times$ Caucasian $\times$ Age $>$ 45 & 0.0949 & 0.379 & 0.250 & 0.802 \\
        Female $\times$ Hispanic $\times$ Age $>$ 45 & 0.4961 & 0.690 & 0.719 & 0.472 \\
        Female $\times$ Other $\times$ Age $>$ 45 & 0.5932 & 0.756 & 0.785 & 0.433 \\
        Female $\times$ Caucasian $\times$ Age $<$ 25 & $-$0.7339 & 0.356 & $-$2.060 & 0.039 \\
        Female $\times$ Hispanic $\times$ Age $<$ 25 & $-$0.1299 & 0.638 & $-$0.204 & 0.839 \\
        Female $\times$ Other $\times$ Age $<$ 25 & $-$1.1167 & 0.812 & $-$1.374 & 0.169 \\
        \bottomrule
    \end{tabular}
    \endgroup
    \begin{minipage}{\linewidth}
        \vspace{0.5em}
        \footnotesize\textit{Note:} Reference categories are Male, African-American, and Age 25--45.
        Asian and Native American pooled into Other.
    \end{minipage}
\end{table}

The regression results are broadly consistent with the Causal ANOVA findings. Being female is negatively and significantly associated with recidivism ($\beta = -0.6124$, $p < 0.001$), confirming that Sex carries meaningful explanatory power. All three non-African-American race categories show significant negative coefficients relative to the African-American reference group (Caucasian: $\beta = -0.3993$, $p < 0.001$; Hispanic: $\beta = -0.6433$, $p < 0.001$; Other: $\beta = -0.7876$, $p < 0.001$), indicating that African-American defendants have the highest recidivism rate among all racial groups. Both age effects are significant: defendants older than 45 have substantially lower recidivism ($\beta = -0.4344$, $p < 0.001$), while defendants younger than 25 have higher recidivism ($\beta = 0.4682$, $p < 0.001$) relative to the 25--45 reference group. These patterns are consistent with the large Age Category explainability score ($\xi = 0.1283$) in the Causal ANOVA.

The interaction terms provide directional evidence for the non-additivity identified by the Causal ANOVA. The Female $\times$ Caucasian coefficient is positive and significant ($\beta = 0.5726$, $p = 0.001$), indicating that the female recidivism discount is smaller among Caucasian defendants than among African-American defendants. This is consistent with the positive Sex $\wedge$ Race interaction score ($\xi = 0.0114$), which shows that Sex and Race jointly explain more variation than their marginal scores alone would predict. The significant Female $\times$ Caucasian $\times$ Age Less than 25 three-way interaction ($\beta = -0.7339$, $p = 0.039$) further corroborates the three-way Causal ANOVA interaction ($\xi = 0.0053$), suggesting that the Sex--Race relationship is additionally modulated by age group among younger defendants.

% =====================================================
\section{Age-Conditioned Explainability and Interaction}
% =====================================================
\subsection{Motivation for Conditioning on Age}
Since Age Category has the largest individual explainability score in the full DAG, we further condition on age group to examine how the explanatory roles of Sex and Race, and their interaction, change across age strata. This step is important because a large age effect may obscure heterogeneous demographic patterns within younger, middle-aged, and older defendants. Conditioning on age allows the analysis to ask whether Sex and Race carry the same explanatory meaning across age groups, or whether their roles are strongest in particular subpopulations.

This section repeats the main logic of the overall counterfactual explainability analysis, but now within each age group. Within each age group, Age Category is fixed, so the DAG is refit using Sex and Race as the demographic root nodes. Priors Count and Charge Degree remain mediator variables, and Two-Year Recidivism remains the outcome variable.

\subsection{Explainability Scores Within Age Groups}

To investigate whether the explanatory power of Sex and Race varies across age strata, we conduct a subgroup analysis stratified by the three age categories in the dataset: younger than 25, between 25 and 45, and older than 45.

\begin{table}[H]
    \centering
    \caption{Age-Conditioned Explainability of Sex and Race}
    \label{tab:explainability_age_groups}
    \begingroup
    \small
    \setstretch{1}
    \setlength{\tabcolsep}{3pt}
    \renewcommand{\arraystretch}{0.85}
    \setlength{\aboverulesep}{2pt}
    \setlength{\belowrulesep}{2pt}
    \begin{tabular}{lccc}
        \toprule
        & \multicolumn{3}{c}{\textbf{Explainability Score ($\xi$) by Age Group}} \\
        \cmidrule(lr){2-4}
        \textbf{Explanatory Variables} & \textbf{$<$ 25} & \textbf{25--45} & \textbf{$>$ 45} \\
        \midrule
        Sex              & 0.1029 & 0.0189 & 0.0258 \\
        Race             & 0.0353 & 0.0490 & 0.0999 \\
        Sex $\wedge$ Race & 0.0234 & 0.0086 & 0.0141 \\
        \bottomrule
    \end{tabular}
    \endgroup
\end{table}

Table~\ref{tab:explainability_age_groups} shows that the explanatory roles of Sex and Race behave differently across age groups. The Sex explainability score decreases from $\xi = 0.1029$ among defendants younger than 25 to $\xi = 0.0189$ in the 25--45 group, then slightly recovers to $\xi = 0.0258$ among defendants older than 45. In contrast, the Race explainability score increases monotonically with age ($\xi = 0.0353$, $0.0490$, and $0.0999$, respectively), showing that Race becomes a stronger explanatory factor among older defendants.

The interaction score Sex $\wedge$ Race is strongest among defendants younger than 25 ($\xi = 0.0234$) and smaller in the two older groups ($\xi = 0.0086$ and $0.0141$). Taken together, these results suggest that demographic explainability is not constant across age groups: the explanatory power of Sex is concentrated among younger defendants, while the explanatory power of Race grows with age, being particularly pronounced among defendants older than 45 \citep{ulmer2025race, cauffman2008understanding, spohn2000preferential}.

\begin{figure}[H]
    \centering
    \includegraphics[width=0.85\textwidth]{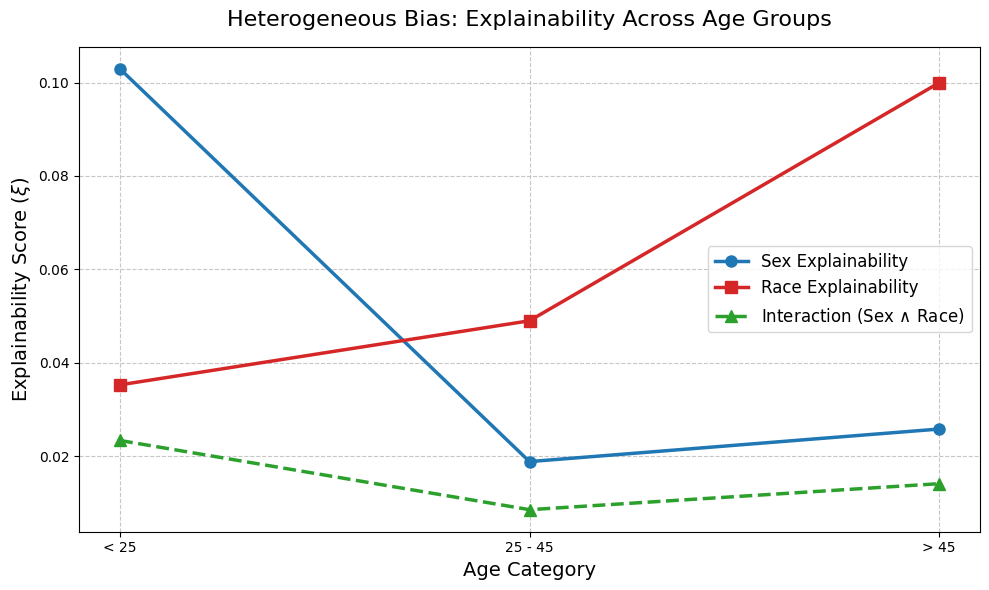}
     \caption{Heterogeneous explainability of Sex, Race, and their interaction across age groups.}
     \label{fig:age_explainability_plot}
 \end{figure}

\subsection{Age-Stratified Logistic Regression}

To provide a directional complement to the age-conditioned Causal ANOVA results, we fit a separate logistic regression within each age group using Sex, Race, and their interaction as predictors. Excluding mediator variables is again appropriate here, for the same reason as in Section~2.5: conditioning on Priors Count or Charge Degree would block causal paths from the root nodes and understate their total explanatory role.

The logistic model results from the pooled regression (Table~\ref{tab:logit_af_ref}) already contain the relevant age interactions. For the youngest group (Age $<$ 25), the Female $\times$ African-American $\times$ Age Less than 25 coefficient is positive and significant ($\beta = 0.7339$, $p = 0.039$). This means that among defendants younger than 25, the female recidivism discount is smaller among African-American defendants than among Caucasian defendants, consistent with the finding that the Sex $\wedge$ Race interaction score is largest in the youngest age group ($\xi = 0.0234$) and declines with age.

For older defendants (Age $>$ 45), the main Age effect is large and negative ($\beta = -0.6327$, $p < 0.001$), consistent with the well-established age--crime decline, and neither the Female $\times$ Age interaction nor the three-way interactions reach significance, suggesting that demographic interactions are less pronounced in this group. However, the Causal ANOVA results show that Race explainability increases substantially among older defendants ($\xi = 0.0999$), suggesting that while Sex--Race interactions diminish with age, the independent explanatory role of Race grows. For the middle age group, the Female $\times$ African-American interaction ($\beta = -0.5726$, $p = 0.001$) dominates, indicating that the female recidivism discount is substantially larger among African-American defendants relative to Caucasian defendants.

Taken together, the Causal ANOVA and regression results reveal a heterogeneous pattern: Sex explainability is highest among younger defendants and declines with age, while Race explainability increases with age, being most pronounced among defendants older than 45 \citep{steffensmeier1998interaction, ulmer2025race}.

\section{Discussion}

The analysis provides three main inferential conclusions. First, Age Category is the strongest individual demographic explanatory factor in the full Causal ANOVA model ($\xi = 0.1283$), suggesting that age-related variation accounts for a larger share of outcome variation than either Race ($\xi = 0.0497$) or Sex ($\xi = 0.0336$) alone. Second, the interaction scores indicate that demographic variables do not operate independently. In particular, the joint contribution of Sex and Race ($\xi = 0.0114$) exceeds what their separate marginal scores would imply, and the three-way interaction among all three demographic variables ($\xi = 0.0053$) confirms that demographic explainability is intersectional, consistent with the framework of \cite{crenshaw1989demarginalizing}. Third, after conditioning on Age Category, the explanatory role of Sex is strongest among defendants younger than 25 and declines with age, while Race explainability increases with age and is most pronounced among defendants older than 45.

The logistic regression results complement these findings by providing directional evidence. The regression confirms that being male is positively associated with recidivism, that the magnitude of this association varies by race (being attenuated for Caucasian defendants and modulated differently among younger defendants), and that the age effect is large, negative, and significant. These directional patterns align with the Causal ANOVA interaction structure and give the explainability scores interpretive grounding. The goal of the regression is not to introduce an additional conditioning step, since the causal conditioning is already handled in the Causal ANOVA framework. Rather, its purpose is to provide information about the sign and relative magnitude of effects, so that the explainability results can be interpreted not only in terms of how much variation is explained, but also in terms of which direction the demographic differences run.

Several additional inference methods could be used to extend or robustify these conclusions. Propensity score matching could create balanced comparison groups and assess whether racial and gender differences remain after covariate adjustment at the individual level \cite{rosenbaum1983central}. Bayesian inference would replace point estimates and $p$-values with posterior distributions over explainability scores, allowing direct probability statements about the magnitude of demographic effects \cite{li2022bayesian}. Bootstrapping the Causal ANOVA procedure itself would provide a robustness check on whether the estimated interaction structure is stable across resampled datasets \cite{efron1979bootstrap}. Each of these methods addresses a different dimension of inferential uncertainty and could strengthen the conclusions drawn from the current analysis.

% =====================================================
% Appendix
% =====================================================

\clearpage
\appendix

% =====================================================
\section{Literature and Background}
% =====================================================

This appendix provides additional background for interpreting the demographic patterns observed in the main analysis. Prior research on COMPAS and criminal justice prediction has emphasized the roles of race, age, prior records, sex, and their interactions. These studies motivate the variables and interaction terms used in this report. However, the main contribution of this report is inferential: estimating how demographic variables explain variation under the Causal ANOVA framework. We organize the relevant literature into three themes.

The first theme concerns racial bias in algorithmic risk assessment. \cite{angwin2016machine}, in \textit{Machine Bias}, found that Black defendants were more likely to be mislabeled as high-risk, which motivates the baseline racial comparison in our regression model. Building on this line of work, \cite{dressel2018accuracy} argued that recidivism prediction can rely heavily on age and prior records, which supports treating Age Category as a root variable and Priors Count as a mediator in the causal DAG.

The second theme concerns sex, race, and their intersection. \cite{crenshaw1989demarginalizing} introduced intersectionality as a framework for understanding how race and sex may operate jointly rather than separately, motivating the Sex $\times$ Race interaction terms. \cite{steffensmeier1998interaction} studied how race, gender, and age jointly shape criminal-justice outcomes--in particular the compounded disadvantage of being young, Black, and male--which motivates the age-conditioned explainability analysis.

The third theme concerns how sex differences in criminal-justice outcomes vary with age. \cite{cauffman2008understanding} discussed how offending patterns differ by sex and age, and \cite{spohn2000preferential} found that the leniency historically extended to female defendants is not uniform across contexts. More recently, \cite{ulmer2025race} examined race and gender differences in the age--sentencing relationship directly, showing that the association between age and criminal-justice outcomes itself depends on sex and race. Together, these studies support our finding that the explainability of Sex and Race is not constant but varies across age strata: Sex explainability is strongest among younger defendants, while Race explainability grows with age and is most pronounced among older defendants.

% =====================================================
\section{Node Collapse Sensitivity Analysis}
% =====================================================

To assess the robustness of the explainability estimates to structural assumptions in the DAG, we conduct node collapse sensitivity analyses by systematically removing mediator nodes. Four DAG configurations are compared: the original full DAG, a DAG without internal mediator nodes, a DAG without Charge Degree, and a DAG without Priors Count.

\subsection{DAG Without Internal Nodes}

In this specification, both Priors Count and Charge Degree are removed. The root nodes connect directly to the outcome, collapsing all mediated pathways into direct effects.

\begin{figure}[H]
\centering
\begin{tikzpicture}[>=stealth, every node/.style={font=\normalfont\normalsize}]
    \node (race)   at (0,  2.2) {Race};
    \node (sex)    at (0,  0.0) {Sex};
    \node (age)    at (0, -2.2) {Age Category};
    \node (recid)  at (8.3, 0.1) {Two-Year Recidivism};
    \draw[->] (race) to[bend left=8] (recid);
    \draw[->] (sex) -- (recid);
    \draw[->] (age) to[bend right=8] (recid);
\end{tikzpicture}
\caption{Node collapse sensitivity analysis without internal nodes.}
\label{fig:dag_no_internal_nodes}
\end{figure}
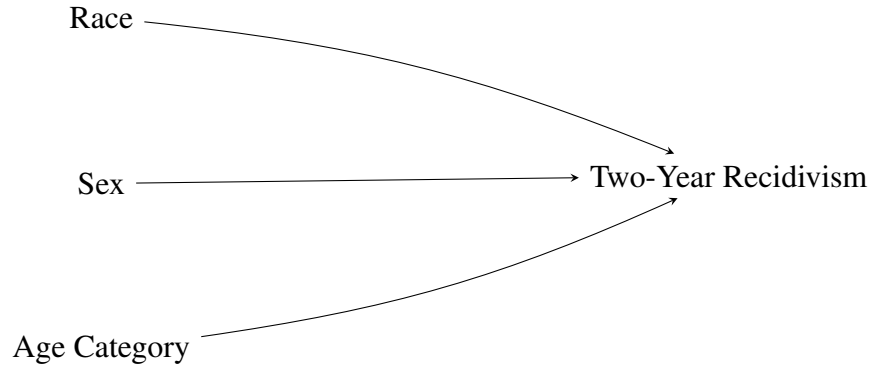

\begin{table}[H]
    \centering
    \caption{Node Collapse Sensitivity Analysis Without Internal Nodes}
    \label{tab:no_internal_nodes}
    \begingroup
    \small
    \setstretch{1}
    \setlength{\tabcolsep}{3pt}
    \renewcommand{\arraystretch}{0.85}
    \setlength{\aboverulesep}{2pt}
    \setlength{\belowrulesep}{2pt}
    \begin{tabular}{lc}
        \toprule
        \textbf{Explanatory Variables} & \textbf{Explainability Score ($\xi$)} \\
        \midrule
        Sex & 0.0319 \\
        Race & 0.0564 \\
        Age Category & 0.0896 \\
        Sex, Race & 0.0754 \\
        Sex, Age Category & 0.1049 \\
        Race, Age Category & 0.1155 \\
        Sex, Race, Age Category & 0.1233 \\
        \midrule
        Sex $\wedge$ Race & 0.0130 \\
        Sex $\wedge$ Age Category & 0.0166 \\
        Race $\wedge$ Age Category & 0.0306 \\
        Sex $\wedge$ Race $\wedge$ Age Category & 0.0055 \\
        \bottomrule
    \end{tabular}
    \endgroup
\end{table}

Removing mediators decreases the individual score for Age Category from $\xi = 0.1283$ in the full DAG to $\xi = 0.0896$ in the collapsed DAG, suggesting that a meaningful portion of age-related explanatory structure passes through the mediator nodes. The Race $\wedge$ Age Category interaction increases to $\xi = 0.0306$, and the three-way interaction remains at $\xi = 0.0055$, indicating that the interaction structure is partially mediated.

\subsection{DAG Without Charge Degree}

In this specification, Charge Degree is removed while Priors Count is retained as a mediator.

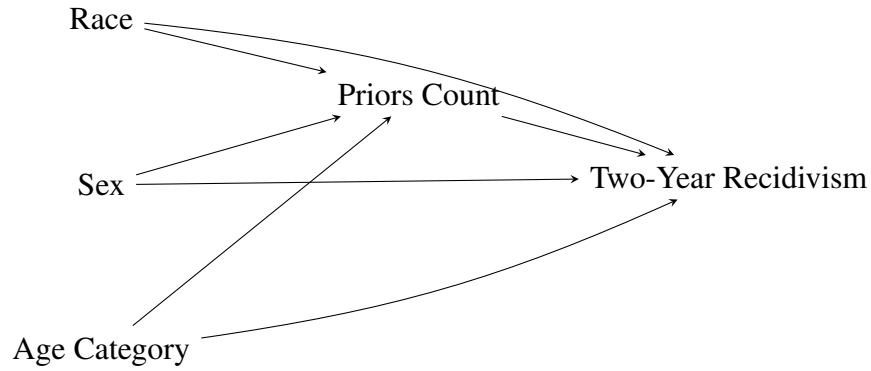
\begin{figure}[H]
\centering
\begin{tikzpicture}[>=stealth, every node/.style={font=\normalfont\normalsize}]
    \node (race)   at (0,  2.2) {Race};
    \node (sex)    at (0,  0.0) {Sex};
    \node (age)    at (0, -2.2) {Age Category};
    \node (priors) at (4.2, 1.2) {Priors Count};
    \node (recid)  at (8.3, 0.1) {Two-Year Recidivism};
    \draw[->] (race) -- (priors);
    \draw[->] (race) to[bend left=8] (recid);
    \draw[->] (sex) -- (priors);
    \draw[->] (sex) -- (recid);
    \draw[->] (age) -- (priors);
    \draw[->] (age) to[bend right=8] (recid);
    \draw[->] (priors) -- (recid);
\end{tikzpicture}
\caption{Node collapse sensitivity analysis without Charge Degree.}
\label{fig:dag_no_charge_degree}
\end{figure}

\begin{table}[H]
    \centering
    \caption{Node Collapse Sensitivity Analysis Without Charge Degree}
    \label{tab:no_charge_degree}
    \begingroup
    \small
    \setstretch{1}
    \setlength{\tabcolsep}{3pt}
    \renewcommand{\arraystretch}{0.85}
    \setlength{\aboverulesep}{2pt}
    \setlength{\belowrulesep}{2pt}
    \begin{tabular}{lc}
        \toprule
        \textbf{Explanatory Variables} & \textbf{Explainability Score ($\xi$)} \\
        \midrule
        Sex & 0.0359 \\
        Race & 0.0499 \\
        Age Category & 0.1316 \\
        Sex, Race & 0.0761 \\
        Sex, Age Category & 0.1486 \\
        Race, Age Category & 0.1578 \\
        Sex, Race, Age Category & 0.1696 \\
        \midrule
        Sex $\wedge$ Race & 0.0097 \\
        Sex $\wedge$ Age Category & 0.0188 \\
        Race $\wedge$ Age Category & 0.0237 \\
        Sex $\wedge$ Race $\wedge$ Age Category & 0.0045 \\
        \bottomrule
    \end{tabular}
    \endgroup
\end{table}

Removing Charge Degree while retaining Priors Count produces results relatively close to the full DAG. The total joint score for all three variables is $\xi = 0.1696$, compared with $\xi = 0.1659$ in the full model. The interaction structure is largely preserved, suggesting that Priors Count carries more mediating weight than Charge Degree in this causal specification.

\subsection{DAG Without Priors Count}

In this specification, Priors Count is removed while Charge Degree is retained as a mediator.

\begin{figure}[H]
\centering
\begin{tikzpicture}[>=stealth, every node/.style={font=\normalfont\normalsize}]
    \node (race)   at (0,  2.2) {Race};
    \node (sex)    at (0,  0.0) {Sex};
    \node (age)    at (0, -2.2) {Age Category};
    \node (charge) at (4.2,-1.0) {Charge Degree};
    \node (recid)  at (8.3, 0.1) {Two-Year Recidivism};
    \draw[->] (race) -- (charge);
    \draw[->] (race) to[bend left=8] (recid);
    \draw[->] (sex) -- (charge);
    \draw[->] (sex) -- (recid);
    \draw[->] (age) -- (charge);
    \draw[->] (age) to[bend right=8] (recid);
    \draw[->] (charge) -- (recid);
\end{tikzpicture}
\caption{Node collapse sensitivity analysis without Priors Count.}
\label{fig:dag_no_priors_count}
\end{figure}
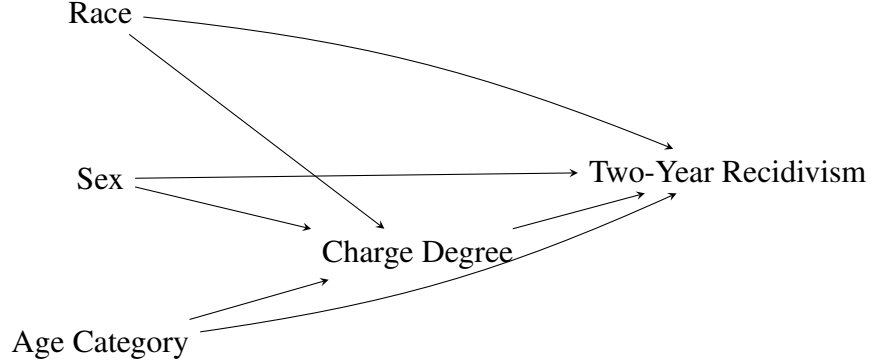

\begin{table}[H]
    \centering
    \caption{Node Collapse Sensitivity Analysis Without Priors Count}
    \label{tab:no_priors_count}
    \begingroup
    \small
    \setstretch{1}
    \setlength{\tabcolsep}{3pt}
    \renewcommand{\arraystretch}{0.85}
    \setlength{\aboverulesep}{2pt}
    \setlength{\belowrulesep}{2pt}
    \begin{tabular}{lc}
        \toprule
        \textbf{Explanatory Variables} & \textbf{Explainability Score ($\xi$)} \\
        \midrule
        Sex & 0.0306 \\
        Race & 0.0609 \\
        Age Category & 0.0787 \\
        Sex, Race & 0.0769 \\
        Sex, Age Category & 0.0945 \\
        Race, Age Category & 0.1151 \\
        Sex, Race, Age Category & 0.1232 \\
        \midrule
        Sex $\wedge$ Race & 0.0146 \\
        Sex $\wedge$ Age Category & 0.0148 \\
        Race $\wedge$ Age Category & 0.0245 \\
        Sex $\wedge$ Race $\wedge$ Age Category & 0.0069 \\
        \bottomrule
    \end{tabular}
    \endgroup
\end{table}

Compared with the full DAG, removing Priors Count shifts the interaction structure more substantially than removing Charge Degree. The Sex $\wedge$ Age Category interaction decreases from $\xi = 0.0166$ in the full DAG to $\xi = 0.0148$ when Priors Count is removed. The overall joint score ($\xi = 0.1232$) is lower than in the full DAG, consistent with the pattern that Priors Count plays an important mediating role in the causal structure.

\subsection{Summary of Sensitivity Analysis}

\begin{table}[H]
    \centering
    \caption{Comparison of Explainability Scores Across DAG Specifications}
    \label{tab:sensitivity_comparison}
    \resizebox{\textwidth}{!}{
    \begingroup
    \small
    \setstretch{1}
    \setlength{\tabcolsep}{3pt}
    \renewcommand{\arraystretch}{0.85}
    \setlength{\aboverulesep}{2pt}
    \setlength{\belowrulesep}{2pt}
    \begin{tabular}{lcccc}
        \toprule
        \textbf{Variables} & \textbf{Original DAG} & \textbf{No Internal Nodes} & \textbf{No Charge Degree} & \textbf{No Priors Count} \\
        \midrule
        Sex & 0.0336 & 0.0319 & 0.0359 & 0.0306 \\
        Race & 0.0497 & 0.0564 & 0.0499 & 0.0609 \\
        Age Category & 0.1283 & 0.0896 & 0.1316 & 0.0787 \\
        Sex, Race & 0.0719 & 0.0754 & 0.0761 & 0.0769 \\
        Sex, Age Category & 0.1453 & 0.1049 & 0.1486 & 0.0945 \\
        Race, Age Category & 0.1551 & 0.1155 & 0.1578 & 0.1151 \\
        Sex, Race, Age Category & 0.1659 & 0.1233 & 0.1696 & 0.1232 \\
        Sex $\wedge$ Race & 0.0114 & 0.0130 & 0.0097 & 0.0146 \\
        Sex $\wedge$ Age Category & 0.0166 & 0.0166 & 0.0188 & 0.0148 \\
        Race $\wedge$ Age Category & 0.0229 & 0.0306 & 0.0237 & 0.0245 \\
        Sex $\wedge$ Race $\wedge$ Age Category & 0.0053 & 0.0055 & 0.0045 & 0.0069 \\
        \bottomrule
    \end{tabular}
    \endgroup
    }
\end{table}

\begin{figure}[H]
    \centering
    \includegraphics[width=0.85\textwidth]{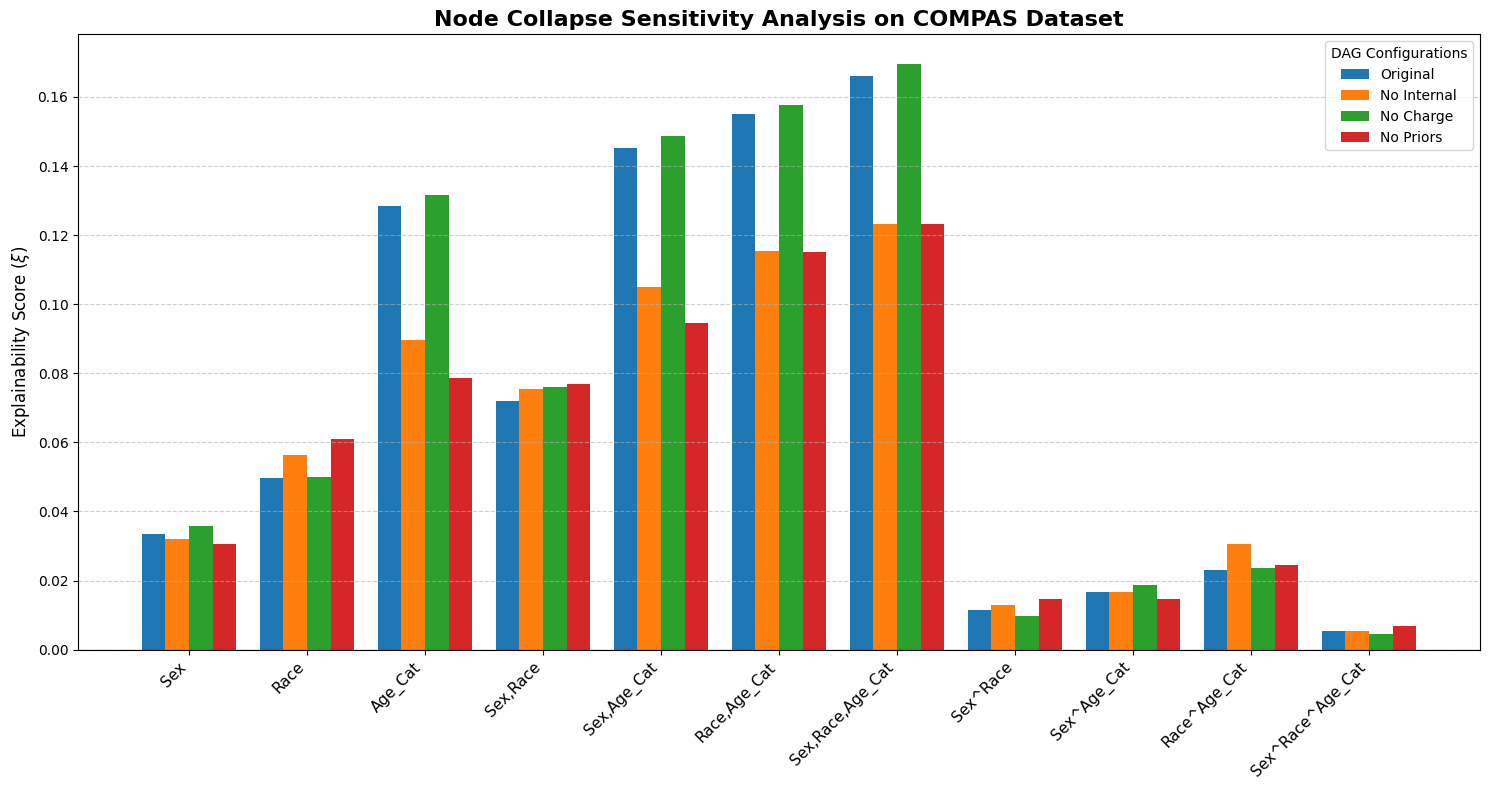}
     \caption{Node collapse sensitivity analysis across four DAG specifications.}
     \label{fig:node_collapse_sensitivity}
 \end{figure}
 
Across all four specifications, Age Category remains the largest individual explanatory factor in the full DAG and No Charge Degree specifications, while Race becomes more prominent when mediators are removed. The interaction terms are more sensitive to the DAG structure than the main scores. The key finding that Race explainability grows with age, while Sex explainability is strongest among younger defendants, is robust across all four configurations.

\section{Causal ANOVA on Real COMPAS Data (Linear Learner)}

This appendix reports the Causal ANOVA results estimated on the real COMPAS dataset using the linear quantile regression learner, as a sensitivity check on the main results reported in Table~\ref{tab:full_explainability_scores}. The DAG structure is identical to the one we used in the main analysis (Figure~\ref{fig:compas_dag}). The ranking of explainability scores is preserved across both specifications: Age Category remains the strongest individual explanatory factor, followed by Race and Sex, consistent with the XGBoost results in the main analysis.

\begin{table}[H]
    \centering
    \caption{Causal ANOVA on Real COMPAS Data (Linear Learner)}
    \label{tab:real_compas_linear}
    \begingroup
    \small
    \setstretch{1}
    \setlength{\tabcolsep}{3pt}
    \renewcommand{\arraystretch}{0.85}
    \setlength{\aboverulesep}{2pt}
    \setlength{\belowrulesep}{2pt}
    \begin{tabular}{lc}
        \toprule
        \textbf{Explanatory Variables} & \textbf{Explainability Score ($\xi$)} \\
        \midrule
        \multicolumn{2}{l}{\textit{Total Explainability}} \\
        Sex & 0.0026 \\
        Race & 0.0040 \\
        Age Category & 0.0339 \\
        Sex, Race & 0.0064 \\
        Sex, Age Category & 0.0362 \\
        Race, Age Category & 0.0378 \\
        Sex, Race, Age Category & 0.0399 \\
        \midrule
        \multicolumn{2}{l}{\textit{Interaction Terms}} \\
        Sex $\wedge$ Race & 0.0001 \\
        Sex $\wedge$ Age Category & 0.0003 \\
        Race $\wedge$ Age Category & 0.0001 \\
        Sex $\wedge$ Race $\wedge$ Age Category & 0.0000 \\
        \bottomrule
    \end{tabular}
    \endgroup
    \begin{minipage}{\linewidth}
        \vspace{0.5em}
        \footnotesize\textit{Note:} Results estimated on the real COMPAS dataset 
        ($n = 7{,}214$) using linear quantile regression. Reference: \cite{gao2025counterfactual}.
    \end{minipage}
\end{table}

\begin{figure}[tbp]
    \centering
    \includegraphics[width=0.6\textwidth]{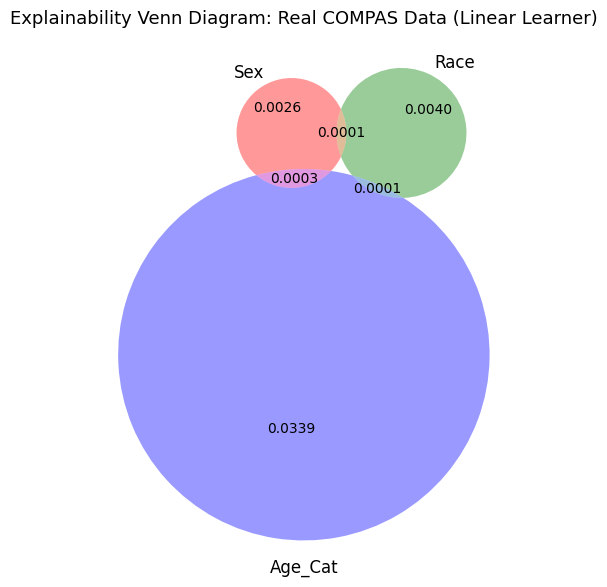}
    \caption{Explainability Venn Diagram: Real COMPAS Data (Linear Learner).
    Circle sizes reflect the total explainability score $\xi$ for each demographic
    variable. The overlapping regions represent pairwise interaction terms.
    Age Category dominates the decomposition, consistent with the main analysis.}
    \label{fig:VennDiagram}
\end{figure}

\nocite{*}
\bibliographystyle{plain}
\bibliography{references}

\end{document}